\documentclass[runningheads]{llncs}
\usepackage[T1]{fontenc}
\usepackage{graphicx,verbatim}
\usepackage{graphicx}
\usepackage{amsmath}
\usepackage{amssymb}
\usepackage{booktabs}
\usepackage{pifont}
\usepackage{colortbl}
\usepackage{algorithm}
\usepackage{algorithmic}
\usepackage{bbding}
\usepackage{soul}
\usepackage{bibunits}
\usepackage{bm}
\usepackage{array}
\usepackage[dvipsnames]{xcolor} 
\usepackage{pdflscape}
\usepackage{makecell}
\usepackage{framed,multirow}
\usepackage{threeparttable}
\usepackage{amssymb}
\usepackage{pifont}
\usepackage{algorithm}
\usepackage{listings}
\usepackage{pythonhighlight}
\usepackage{float}
\usepackage{mathtools}
\usepackage{cuted}
\makeatletter
\newcommand\footnoteref[1]{\protected@xdef\@thefnmark{\ref{#1}}\@footnotemark}
\makeatother

\newcolumntype{P}[1]{>{\centering\arraybackslash}p{#1}}
\newlength\savewidth
\newcommand{\eg}{\mbox{e.g.,\ }}
\newcommand{\etal}{\mbox{et al.}}

\def\arrvline{\hfil\kern\arraycolsep\vline\kern-\arraycolsep\hfilneg}
\definecolor{mygray}{gray}{.9}
\definecolor{Highlight}{HTML}{39b54a}  

\newcolumntype{x}[1]{>{\centering\arraybackslash}p{#1pt}}
\newcolumntype{z}[1]{>{\raggedright\arraybackslash}p{#1pt}}

\definecolor{citecolor}{HTML}{0071BC}
\definecolor{linkcolor}{HTML}{ED1C24}
\usepackage[pagebackref,breaklinks,colorlinks,citecolor=citecolor]{hyperref}

\usepackage[T1]{fontenc}

\definecolor{iblue}{rgb}{0.06, 0.75, 1.0}
\definecolor{ired}{RGB}{255 69 0}

\begin{document}
\title{Exploring the Role of Autoregressive Models in Organ Motion Prediction for Radiotherapy}
\title{Patient-Specific Autoregressive Models for Organ Motion Prediction in Radiotherapy}
%

\newcommand{\equalcontribmark}{\textsuperscript{*}}
\newcommand{\correspondencemark}{\textsuperscript{\dag}}

\newcommand{\equalcontribtext}{\begingroup
\renewcommand\thefootnote{*}
\footnotetext{These authors contributed equally to this work.}
\endgroup}

\newcommand{\correspondencetext}{\begingroup
\renewcommand\thefootnote{\dag}
\footnotetext{Correspondence to: Xiaofeng Yang (\href{mailto:xyang43@emory.edu}{\texttt{xyang43@emory.edu}})}
\endgroup}

\author{
Yuxiang Lai\equalcontribmark \inst{1} \and 
Jike Zhong\equalcontribmark \inst{2} \and 
Vanessa Su \inst{1} \and 
Xiaofeng Yang\correspondencemark \inst{1,3}
}

\authorrunning{Y. Lai et al.}

\institute{
Emory University \and 
University of Southern California \and 
Georgia Institute of Technology
}

\maketitle              
\equalcontribtext
\correspondencetext
\begin{abstract} 
Radiotherapy often involves a prolonged treatment period. During this time, patients may experience organ motion due to breathing and other physiological factors. Predicting and modeling this motion before treatment is crucial for ensuring precise radiation delivery. 
However, existing pre-treatment organ motion prediction methods primarily rely on deformation analysis using principal component analysis (PCA), which is highly dependent on registration quality and struggles to capture periodic temporal dynamics for motion modeling.
In this paper, we observe that organ motion prediction closely resembles an autoregressive process, a technique widely used in natural language processing (NLP). Autoregressive models predict the next token based on previous inputs, naturally aligning with our objective of predicting future organ motion phases. Building on this insight, we reformulate organ motion prediction as an autoregressive process to better capture patient-specific motion patterns. Specifically, we acquire 4D CT scans for each patient before treatment, with each sequence comprising multiple 3D CT phases. These phases are fed into the autoregressive model to predict future phases based on prior phase motion patterns. We evaluate our method on a real-world test set of 4D CT scans from 50 patients who underwent radiotherapy at our institution and a public dataset containing 4D CT scans from 20 patients (some with multiple scans), totaling over 1,300 3D CT phases. The performance in predicting the motion of the lung and heart surpasses existing benchmarks, demonstrating its effectiveness in capturing motion dynamics from CT images. These results highlight the potential of our method to improve pre-treatment planning in radiotherapy, enabling more precise and adaptive radiation delivery.


\end{abstract}
\section{Introduction}
\begin{figure*}[t]
	\centering
\includegraphics[width=0.9\linewidth]{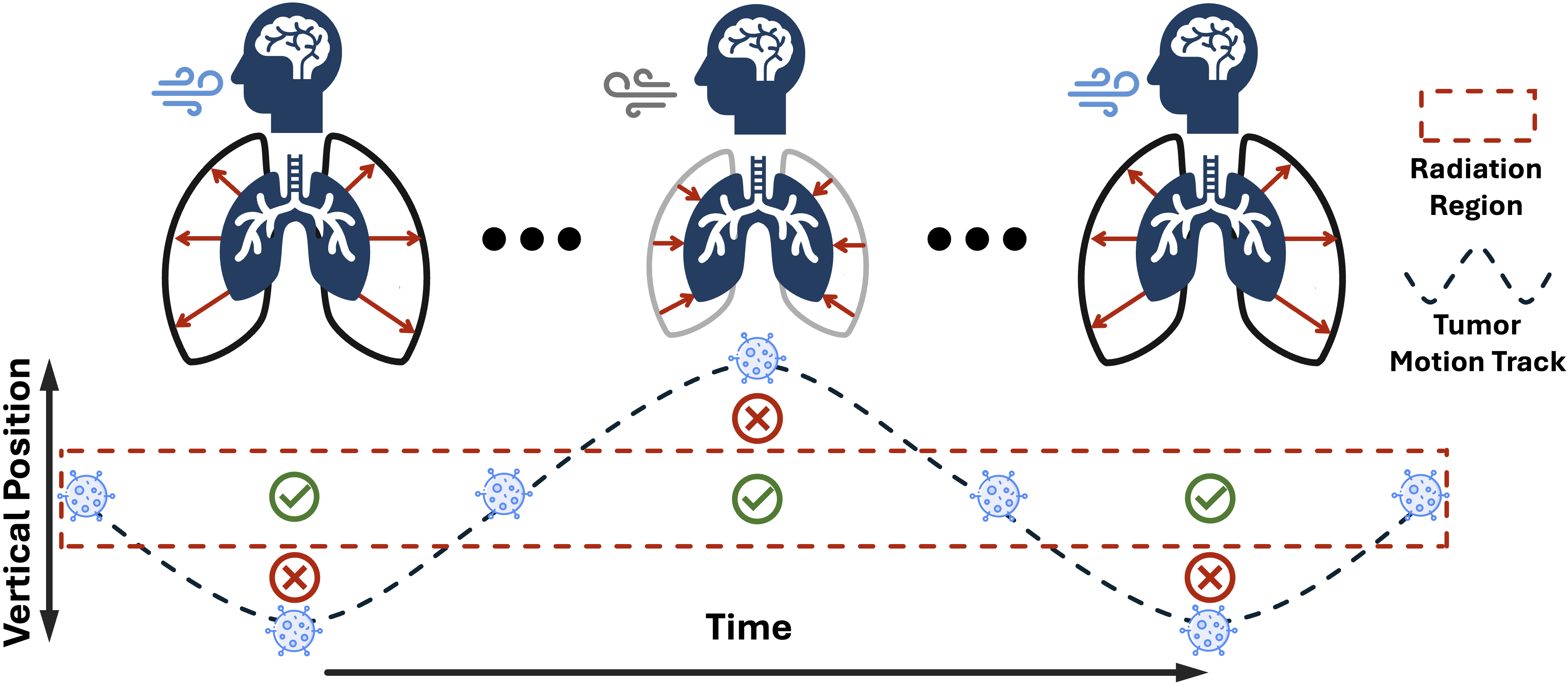}
	\caption{Throughout the extended duration of radiotherapy, patients' natural breathing motion results in the continuous rhythmic expansion and contraction of the lungs. Consequently, the tumor moves up and down in the coronal view (dashed curve). However, current treatment plans often designate the target with a large margin for dose delivery (red rectangle), which accounts for tumor motion but results in unintended radiation doses to surrounding healthy tissues. By modeling and predicting organ motion patterns, dynamic radiation regions can be pre-defined, reducing target margins to enable more precise and adaptive radiation delivery.}
    
    \label{fig_main_idea}
\end{figure*}

 Radiotherapy, a cornerstone of cancer treatment, often spans a period through each treatment for radiation delivery~\cite{chandra2021contemporary,schaue2015opportunities}. Throughout this period, organ motion caused by natural physiological processes such as breathing and cardiac activity pose significant challenges to accurate radiation delivery~\cite{cole2014motion,korreman2012motion}. Even slight shifts in organ position can result in insufficient radiation to the tumor or excessive exposure to surrounding healthy tissues. Therefore, current treatment plans typically use a large target margin to cover most of the tumor's motion range, but this unfortunately leads to greater unintended damage to nearby healthy tissues~\cite{de2019radiotherapy,hubenak2014mechanisms}. Consequently, predicting and modeling organ motion before treatment has become a critical component of modern radiotherapy workflows (\figureautorefname~\ref{fig_main_idea}). The accurate prediction of motion patterns enables clinical teams to deliver more precise and personalized radiation therapy, especially for tumors near critical organs, minimizing damage and maximizing efficacy.

Preliminary studies have explored organ motion prediction in radiotherapy. Before the advent of deep learning, organ motion prediction in radiotherapy primarily relied on statistical methods. 
Principal component analysis (PCA) was widely employed to characterize organ motion and generate deformation vector fields (DVFs) by modeling statistical patterns from population data~\cite{nie2013site,vile2015statistical}. With the rise of deep learning~\cite{tu2023holistic,yu2023multimodal,chen2023federated,zhong2023making,zhang2022learning}, several studies have replaced traditional statistical models with deep learning approaches, such as variational autoencoders (VAEs)~\cite{oord2017neural,van2017neural,lai2023memory} and diffusion models~\cite{ho2020denoising,rombach2022high}, to capture better the statistical patterns modeled by PCA and improve DVF prediction~\cite{pastor2023probabilistic,smolders2024diffusert}. 


Although previous studies have made progress, several key challenges remain:
(i) Ground truth DVFs are typically generated via deformable image registration~\cite{oh2017deformable,fu2020deep}, which is very time-consuming and depends on registration performance.
(ii) Organ motion and anatomical changes vary significantly among patients, and current predictions based highly on the training distribution are unreliable and inaccurate.
(iii) Most existing models are limited to predicting only the next phase of organ motion, failing to capture long-term motion patterns. 
Furthermore, these models typically use only a single phase as input, neglecting patient-specific motion patterns, which limits their applicability and reliability.

\begin{figure*}[t]
	\centering
\includegraphics[width=0.95\linewidth]{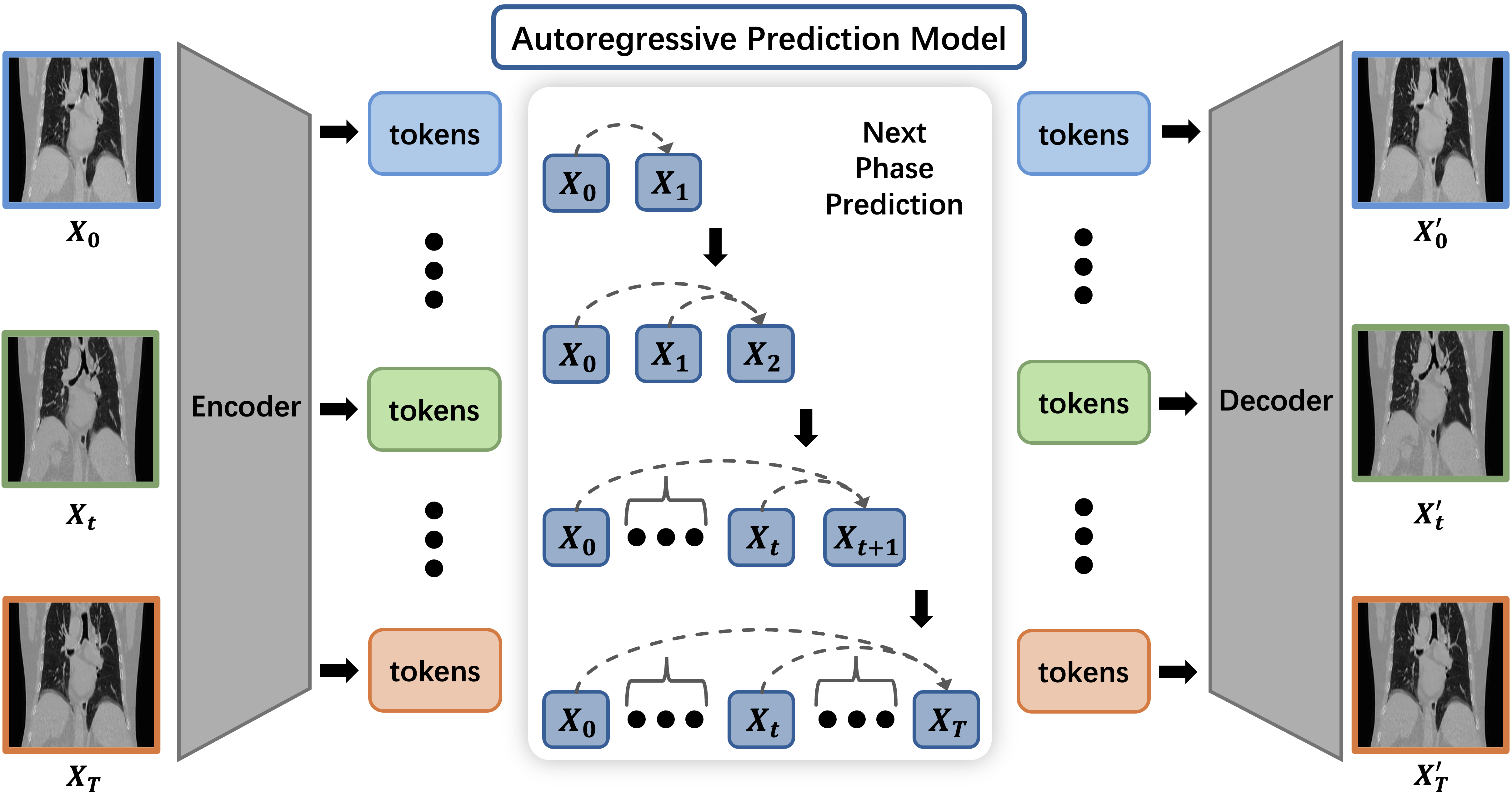}
\caption{\textbf{Auto-RMP.} We first arranges each phase of a 4D CT scan into an input image sequence ($X_{0}$ to $X_{T}$). Then, VQGAN serves as the vision encoder to convert CT images into discrete tokens. The output tokens are arranged into a 1D sequence, which is fed into the Autoregressive Transformer Prediction model. During the autoregressive process, the model predicts the next phase tokens based on the previous phase. Finally, the predicted phase tokens are decoded back into CT phases using the VQGAN decoder.}

    \label{fig_pipeline}
\end{figure*}

In this paper, we propose a method for organ motion prediction for radiotherapy planning. 
Since 4D CT inherently comprises multiple 3D CT phases (\eg 10 3D CT), similar to videos~\cite{underberg2005use,kwong2015f}, we formulate organ motion prediction as an autoregressive process~\cite{bai2024sequential,el2024scalable,yu2021vector,li2025think,li2025cls,li2025towards,lai2025med,li2024eee,li2024vision} (\S\ref{sec:autoregressive}). In this process, the goal is to predict future phases based on the prior phase sequences of a patient. We can capture patient-specific motion patterns by leveraging sequential modeling, improving prediction accuracy.
Moreover, our autoregressive training framework enables direct learning of motion patterns from CT image sequences, removing the dependence on precomputed DVFs. As a result, our proposed \textbf{Auto-RMP} (Autoregressive-Radiotherapy-Motion-Prediction) can predict multiple future CT phases, facilitating more precise and adaptive radiotherapy planning (\figureautorefname~\ref{fig_pipeline}).

We evaluate Auto-RMP on two datasets with different distributions. First, we use a public 4D CT dataset~\cite{hugo2017longitudinal}, which includes 20 patients with locally advanced non-small cell lung cancer. Second, we curate a private 4D CT dataset from our institution, comprising 50 patients with multi-stage lung cancer, lung adenocarcinoma, or lung nodules. On both datasets, Auto-RMP outperforms existing benchmarks in predicting lung and heart motion while achieving high absolute accuracy in terms of IoU and DSC (\tableautorefname~\ref{tab:performance}). These results highlight the potential of Auto-RMP in assisting pre-treatment organ motion prediction.

This performance is attributable to our key observation: Organ motion prediction, with planning and follow-up 4D CTs, can be formulated as an autoregressive problem. The autoregressive model effectively captures the relationships among input CT phases and learns patient-specific motion patterns. We compare \textbf{Auto-RMP} with \textbf{existing methods} to highlight its key advantages. 

\begin{enumerate}

\item \textbf{Eliminating the need for generated DVFs.} Statistical models~\cite{nie2013site,vile2015statistical} and deep learning approaches~\cite{pastor2023probabilistic} heavily rely on precomputed DVFs to learn motion patterns and predict possible deformations. However, this introduces inaccuracies and cannot capture periodic temporal information in 4D CT.

\item  \textbf{Capturing patient-specific motion patterns.} Most existing models rely on single-image inputs and DVFs~\cite{oord2017neural,ho2020denoising}, making their predictions highly dependent on the training data distribution. In contrast, Auto-RMP processes long-sequence CT images, enabling more accurate modeling of patient-specific motion patterns (\tableautorefname~\ref{tab:performance}).

\item \textbf{Predicting multiple future CT phases.} Unlike prior methods that focus on single-step predictions~\cite{pastor2023probabilistic,smolders2024diffusert}, we reformulate organ motion prediction as an autoregressive process. This enables Auto-RMP to generate multi-stage future CT phases sequentially, leveraging long-term temporal dependencies for more precise and realistic motion modeling (\figureautorefname~\ref{fig_longterm}).

\end{enumerate}

\section{Autoregressive Radiotherapy Motion Prediction}\label{sec:autoregressive}

\textbf{Overview.} We first formulate organ motion prediction as an autoregressive process. A 4D CT scan consists of multiple 3D CT acquired at different breathing phases~\cite{hugo2017longitudinal,kwong2015f}, with each phase denoted as $X_t$, where $t \in \mathcal{T} = \{0, 1, ..., T\}$, and $T$ represents the total number of phases. Typically, 4D CT consists of 10 phases, each corresponding to a specific time point within the breathing cycle, allowing for continuous motion tracking of organs. 
Given a sequence of prior CT phases $\{X_0, ..., X_T\}$, our goal is to learn patient-specific motion patterns and predict future phases to support pretreatment radiotherapy planning. This can be achieved by maximizing the following objective:
\begin{align}
P(X_{1}, X_{2}, \dots, X_{T}) = \prod_{t=1}^{T} p_\theta(X_t' \mid X_0, X_1, \dots, X_{t-1})
\end{align}
where $X_t'$ represents the predicted phase at time $t$, and $p_\theta$ is a model parameterized by $\theta$ that takes a sequence of prior CT phases as input to predict the next phase. During inference, we can then view the patient's prior CT scan history $\{X_0, \dots, X_{L}\}$ for $(L<T)$  as the prefix and predict motion patterns of future phases $\{X_{L+1}',\dots, X_{T}'\}$ by maximizing the conditional likelihood $P(X'_{L+1}, \dots, X'_{T} \mid X_0, \dots, X_L)$:
\begin{align}
\max\limits_{\theta} \prod_{t=L+1}^{T} p_\theta(X_t' \mid X_0,\dots, X_{t-1})
\end{align}
This formulation naturally aligns with sequence generation tasks, allowing us to leverage the autoregressive modeling techniques. With this formulation, our method is illustrated as follows.

\subsection{Pipeline}

\smallskip\noindent\textbf{Pre-processing.}
We first segment the main organs in each phase of the 4D CT scan. To achieve this, we employ nnUNet~\cite{isensee2021nnu}, pre-trained with TotalSegmentator~\cite{wasserthal2023totalsegmentator}, for automatic segmentation of the lungs and heart in CT images. These segmentations provide ground-truth labels for organ shape and position, which are later used as auxiliary inputs and for evaluating predictions.

\smallskip\noindent\textbf{CT tokenization.} 
To apply autoregressive transformer models to images, some previous methods typically divide images into patches and flatten them into one-dimensional sequences of tokens~\cite{dosovitskiy2020image}. In Auto-RMP, we adopt an alternative tokenization strategy by encoding images into a discrete codebook representation using vector quantization (VQ)~\cite{van2017neural,lai2023memory}. This enables us to represent CT images as a sequence of learnable tokens. 

As shown in~\figureautorefname~\ref{fig_pipeline}, we generate token sequences for an autoregressive prediction model and decode the model output back into CT images. Specifically, VQGAN~\cite{esser2021taming} is employed in this pipeline. VQGAN is a generative model comprising an encoder, a decoder, and a quantization codebook. Both the encoder and decoder are built using convolutional neural networks (CNNs). The encoder extracts spatial latent features from input images through multiple downsampling layers, which are then quantized into discrete token sequences using the quantization codebook. The decoder reconstructs CT images from the output token sequences via aligned upsampling layers.


\smallskip\noindent\textbf{Sequence motion modeling of CT phases.} 
After all input CT phases are encoded into token sequences, these tokens are processed using an autoregressive model. The model is built upon unidirectional causal self-attention transformer layers, similar to those used in large language models~\cite{brown2020language,vaswani2017attention}. This architecture enables the model to predict the token sequence for the next CT phase by conditioning on the token sequences of previously observed phases. By leveraging sequential dependencies, the model effectively captures temporal patterns in organ motion, ensuring more consistent and accurate phase generation. 

\subsection{Implementation Details}  
We use the nnUNet model from the TotalSegmentator toolkit~\cite{wasserthal2023totalsegmentator} to extract organ masks from CT images. This UNet-based~\cite{ronneberger2015u} framework automatically adjusts hyperparameters based on dataset characteristics~\cite{li2024abdomenatlas,lai2024pixel,chen2024analyzing}. The VQGAN module employs a downsampling factor of \( f = 16 \) and a codebook size of 8192, generating a \(16 \times 16 = 256\) token grid for each \(256 \times 256\) image. We adopt pre-trained parameters from Yutong~\etal~\cite{bai2024sequential}.  

Each CT image in the 4D CT sequence is first tokenized into 256 tokens, which are then flattened into a 1D sequence. Auto-RMP follows an autoregressive modeling approach and adopts the Transformer architecture of LLaMA~\cite{touvron2023llama}, a widely used open-source language model with publicly available implementations. The model supports a context length of 4096 tokens, allowing it to process up to 16 images within the VQGAN tokenizer framework. 

\section{Experiment \& Result}

\subsection{Experiment}

\smallskip\noindent\textbf{Dataset.}  
Our study utilizes two 4D CT datasets. The first dataset, provided by Hugo~\etal~\cite{hugo2017longitudinal}, is publicly available and includes 4D CT scans from 20 patients diagnosed with locally advanced non-small cell lung cancer. The images were acquired using a 16-slice helical CT scanner (Brilliance Big Bore, Philips Medical Systems, Andover, MA) with respiration-correlated CT imaging. Each scan comprises 10 breathing phases (0\% to 90\%, phase-based binning) with a slice thickness of 3 mm.
The second dataset was acquired at our institute. Each 4D CT scan, similar to the first dataset, consists of 10 sets of 3D CT images corresponding to different breathing phases. All scans were performed using a Siemens SOMATOM Definition AS scanner at 120 kV, following the standard Siemens Lung 4D CT protocol. The imaging system was configured with Syngo CT VA48A software, a pitch of 0.8, and a reconstruction kernel of Bf37. The image resolution is \( 512 \times 512 \times (133\text{-}168) \), with a voxel spacing of \( 0.9756 \times 0.9756 \times 2 \) mm\(^3\) along the axial, coronal, and sagittal axes.

 \begin{table*}[t]
    \centering
    \scriptsize
    \caption{\textbf{Performance on Next-Phase Motion Prediction.} Auto-RMP significantly outperforms previous methods and achieves high accuracy. We evaluate performance by comparing the segmentation masks of organs between the predicted and ground-truth CT phases. Higher IoU and NSD values indicate more precise overall predictions, while lower SD and HD values suggest more accurate boundary alignment.}

\begin{tabular}{P{0.09\linewidth}|P{0.09\linewidth}P{0.24\linewidth}|P{0.13\linewidth}P{0.13\linewidth}P{0.13\linewidth}P{0.13\linewidth}}
        \toprule
        Data & Organ & Method & IoU (\%) $\uparrow$ & DSC (\%) $\uparrow$ & SD (mm) $\downarrow$ & HD (mm) $\downarrow$ \\
        \midrule 
       \multirow{6}{*}{Public} & \multirow{3}{*}{Lung} & DAM~\cite{pastor2023probabilistic} & 80.03 & 85.53 & 10.80 & 28.61\\
       & & DiffuseRT~\cite{smolders2024diffusert} & 81.16 & 86.95 & 9.13 & 27.75\\
       & & Auto-RMP (Ours)& \cellcolor{ired!20} 90.75 & \cellcolor{ired!20}95.15 & \cellcolor{ired!20}7.45 & \cellcolor{ired!20}21.66\\

        \cmidrule{2-7}
        
        & \multirow{3}{*}{Heart} & DAM~\cite{pastor2023probabilistic} & 78.65 & 82.73 & 11.57 & 32.17\\
       & & DiffuseRT~\cite{smolders2024diffusert} & 80.37 & 85.02 & 10.16 & 29.44\\
       & & Auto-RMP (Ours)& \cellcolor{ired!20}88.43& \cellcolor{ired!20}93.85 & \cellcolor{ired!20}6.88 & \cellcolor{ired!20}22.04\\
        
        \midrule 
       \multirow{6}{*}{Private} & \multirow{3}{*}{Lung} & DAM~\cite{pastor2023probabilistic} & 79.68 & 83.18 & 11.48 & 31.43\\
       & & DiffuseRT~\cite{smolders2024diffusert} & 82.47 & 85.39 & 10.89 & 27.84\\
       & & Auto-RMP (Ours)&  \cellcolor{ired!20}91.88 & \cellcolor{ired!20}95.74 & \cellcolor{ired!20}6.57 & \cellcolor{ired!20}20.73\\

        \cmidrule{2-7}
        
        & \multirow{3}{*}{Heart} & DAM~\cite{pastor2023probabilistic} & 76.77 & 81.00 & 13.29 & 34.83\\
       & & DiffuseRT~\cite{smolders2024diffusert} & 79.03 & 83.46 & 12.60 & 30.74\\
       & & Auto-RMP (Ours)& \cellcolor{ired!20}87.83 & \cellcolor{ired!20}93.24 & \cellcolor{ired!20}7.38 & \cellcolor{ired!20}23.97\\
        
        \bottomrule
    \end{tabular}
    \begin{tablenotes}
        \item Public: 80 4D CT scans, 800 3D CT scans; Private: 50 4D CT scans, 500 3D CT scans.
        \item IoU - intersection over union; DSC - dice similarity coefficient. 
        \item SD - surface distance; HD - Hausdorff distance.
    \end{tablenotes}
    \label{tab:performance}
\end{table*}

\smallskip\noindent\textbf{Evaluation.} Since our focus is on predicting organ motion in radiotherapy, we primarily evaluate the predicted organ locations and shapes. Therefore, we do not assess the overall quality of the generated predictions but instead focus on the segmentation masks of each organ. Specifically, we measure the alignment between the predicted and ground-truth masks using the following evaluation metrics: IoU (Intersection over Union), DSC (Dice Similarity Coefficient), SD (Surface Distance), and HD (Hausdorff Distance). Also, we combine the CT image and segmentation mask as input of the model.

 \subsection{Results}

\smallskip\noindent\textbf{Next-Phase Motion Prediction.}  
We evaluate Auto-RMP against the latest deep learning-based methods for organ motion prediction, including DAM~\cite{pastor2023probabilistic} and DiffuseRT~\cite{smolders2024diffusert}. As shown in \tableautorefname~\ref{tab:performance}, Auto-RMP outperforms existing approaches, achieving an IoU of 90.75\% and a DSC of 95.15\% for lung motion prediction on the public dataset. Additionally, Auto-RMP demonstrates superior performance in heart motion prediction, another critical aspect of chest 4D CT, with an IoU of 88.43\% and a DSC of 93.85\%.

We further analyze the underlying reasons for the significant performance gap between autoregressive models and traditional generative models. Traditional generative models typically rely on limited input conditions, often a single image, which constrains their ability to capture long-term motion patterns such as the breathing cycle. As a result, their predictions tend to be less accurate for continuous motion.
In contrast, Auto-RMP conditions its predictions on a sequence of past inputs, enabling it to capture patient-specific motion patterns more accurately. Additionally, it incorporates its own predictions into the input sequence for future steps, creating a feedback mechanism that helps maintain temporal consistency. This approach aligns more naturally with the problem setting of organ motion prediction, as defined in \S\ref{sec:autoregressive}.

\begin{figure*}[t]
	\centering
\includegraphics[width=0.95\linewidth]{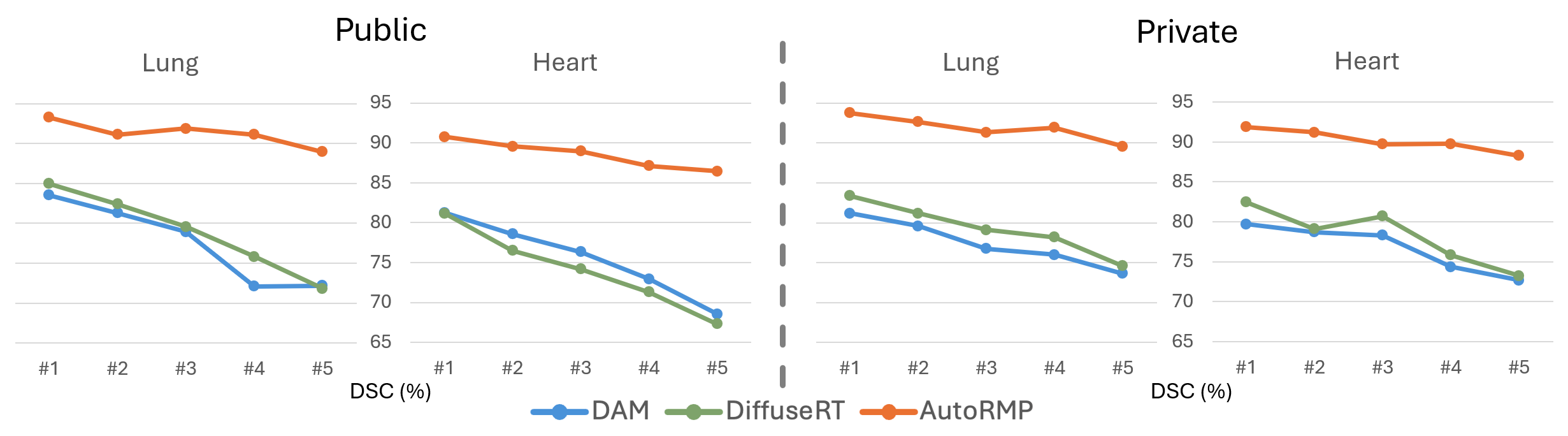}
\caption{\textbf{Long-term motion prediction.} We evaluate the ability of models to perform long-term motion prediction. In this setting, the model is given the first five phases of the 4D CT scan and must predict the next five phases to assess its capability for learning long-term motion patterns. Auto-RMP demonstrates strong robustness in long-term motion prediction, generating smooth and consistent motion predictions.}


    \label{fig_longterm}
\end{figure*}

\smallskip\noindent\textbf{Long-term Motion Prediction.}  
The key difference between traditional generative models and autoregressive models is that the latter can generate multiple motion phases, whereas traditional generative models typically produce only a single phase without time consistency. To assess their effectiveness in long-term motion prediction, we evaluate Auto-RMP alongside two other generative models.  
In our experiment, we provide the first five phases of a 4D CT scan as input and instruct the model to predict the subsequent five phases, which are then compared against the ground-truth 4D CT phases (\figureautorefname~\ref{fig_longterm}). We observe that the performance of common generative models decreases significantly as the number of predicted phases increases. In contrast, the autoregressive model maintains stable performance, achieving over 85\% DSC even in the final predicted phase.  

This also highlights a key limitation of conventional generative models: they are often restricted to generating predictions for a single phase. When predicting subsequent phases, they must rely solely on the single previous prediction as input conditions, leading to error propagation across multiple phases. Auto-RMP demonstrates strong robustness in long-term motion prediction, as its autoregressive nature enables it to retain past input information and previously predicted results. This allows Auto-RMP to generate smooth and consistent motion trajectories, effectively reducing errors over multiple phases.  

\begin{table*}[t]
    \centering
    \scriptsize
    \caption{\textbf{Ablation Study.} We evaluate Auto-RMP under three input conditions: using only CT images, using only segmentation masks (representing organ locations and shapes), and combining both CT images and segmentation masks.}

\begin{tabular}{P{0.09\linewidth}|P{0.09\linewidth}P{0.24\linewidth}|P{0.13\linewidth}P{0.13\linewidth}P{0.13\linewidth}P{0.13\linewidth}}
        \toprule
        Data & Organ & Input & IoU (\%) $\uparrow$ & DSC (\%) $\uparrow$ & SD (mm) $\downarrow$ & HD (mm) $\downarrow$ \\
        \midrule 
       \multirow{6}{*}{Mixed} & \multirow{3}{*}{Lung} & CT Only & 85.83 & 88.57 & 11.98 & 24.30\\
       & & Mask Only & 87.60 & 91.88 & 9.48 & 21.43\\
       & &  CT + Mask & \cellcolor{ired!20}91.23 & \cellcolor{ired!20}94.87 & \cellcolor{ired!20}7.61 & \cellcolor{ired!20}20.92\\

        \cmidrule{2-7}
        
        & \multirow{3}{*}{Heart} & CT Only & 83.71 & 86.83 & 12.77 & 29.67\\
       & & Mask Only & 84.25 & 88.12 & 10.24 & 27.90\\
       & & CT + Mask & \cellcolor{ired!20}87.38 & \cellcolor{ired!20}91.11 & \cellcolor{ired!20}7.94 & \cellcolor{ired!20}24.36\\
        \bottomrule
    \end{tabular}

    \begin{tablenotes}
        \item Mixed: 20 4D CT scans of public dataset, 20 4D CT scans of our private dataset
        \item IoU - intersection over union; DSC - dice similarity coefficient. 
        \item SD - surface distance; HD - Hausdorff distance.
    \end{tablenotes}
    
    \label{tab:ablation}
\end{table*}

\smallskip\noindent\textbf{Ablation Study.}  
As shown in \tableautorefname~\ref{tab:ablation}, using only CT images as input makes it more difficult to capture organ motion regions within the entire scan. When using only segmentation masks, the model better captures organ boundaries and motion-specific regions. However, this approach loses important anatomical information that can only be extracted from CT images. Therefore, we choose to combine CT images and segmentation masks as input for Auto-RMP, enabling the model to leverage complementary features from both modalities and learn patient-specific motion patterns more effectively.

\section{Conclusion \& Discussion}  

In this paper, we formulate organ motion prediction in radiotherapy as an autoregressive process and introduce Auto-RMP, which effectively predicts future organ motion phases. We offer a promising direction for improving radiotherapy motion management through more accurate organ motion modeling.
Although Auto-RMP demonstrates strong performance, its evaluation is currently limited to single-session 4D CT, representing only a partial view of real clinical scenarios. In practice, organ and lesion morphology evolve over multiple radiotherapy sessions spanning weeks or months, posing additional challenges.
For future work, we are collecting longitudinal radiotherapy data from liver cancer patients to study multi-session motion prediction. Extending Auto-RMP to model long-term dynamics will further enhance the precision, adaptability, and clinical relevance of radiotherapy planning.




    


\bibliographystyle{splncs04}
\bibliography{ref}



\end{document}